\documentclass[reprint,aip,amsmath,amssymb,amsfonts,showkeys]{revtex4-2}
\usepackage[dvipdfmx]{graphicx}
\usepackage{dcolumn}
\usepackage{bm}
\usepackage{txfonts}
\usepackage{multirow}
\usepackage{color}

\begin{document}

\title{Mechanically-exfoliated low-layered [Ca$_2$CoO$_3$]$_{0.62}$[CoO$_2$]:
A single-crystalline p-type transparent conducting oxide}

\author{Reiji~Okada}
\thanks{Authors to whom correspondence should be addressed: [Reiji~Okada, 6222508@ed.tus.ac.jp; Yoshiki~J.~Sato, yoshiki\_sato@rs.tus.ac.jp; Ryuji~Okazaki, okazaki@rs.tus.ac.jp]}
\author{Hiroto~Isomura}
\author{Yoshiki~J.~Sato}
\author{Ryuji~Okazaki}
\author{Masayuki~Inoue}
\author{Shinya~Yoshioka}
\affiliation{Department of Physics and Astronomy, Tokyo University of Science, Noda 278-8510, Japan}

\begin{abstract}

Transparent conducting oxides (TCOs) are essential components of optoelectronic devices 
and
various materials have been explored for highly efficient TCOs having a combination of high transmittance and low sheet resistance.
Here, 
we focus on a misfit thermoelectric oxide [Ca$_2$CoO$_3$]$_{0.62}$[CoO$_2$] and 
fabricate the transparent low-layered crystals by a mechanical tape-peeling method using the single-crystalline samples. 
From the transmittance measurement, 
we find that the thickness of low-layered samples is several orders of hundred nanometers, 
which is comparable with the estimation from 
the scanning electron microscopy images.
Compared to the previous results on the polycrystalline and $c$-axis oriented transparent films,
the electrical resistivity is reduced 
owing to the single-crystalline nature.
The figure of merit for the transparent conducting materials  in the present low-layered samples is then evaluated
to be
higher than the values  in the previous reports.
The present results on 
the low-layered single-crystalline [Ca$_2$CoO$_3$]$_{0.62}$[CoO$_2$]
may offer
a unique class of multi-functional transparent thermoelectric oxides.

\end{abstract}

\maketitle

Transparent conducting oxides (TCOs) show good electrical conductivity and optical transparency in the visible range, 
and are widely adopted as components of optoelectronic devices such as liquid crystal displays and 
touch screens \cite{Minami2005,n-type2012,n-type2016}. 
Among such TCOs, n-type materials including 
indium oxides \cite{InO1986,InO1994,InO1995,InO1998,InO2001}, 
zinc oxides \cite{Yamada1997}, 
tin oxides \cite{Wang2009},
and titanium oxides \cite{Hitosugi2010} 
are known as highly efficient TCOs and have been often used as those components.
On the other hand, 
p-type TCOs are less used due to their low carrier mobility \cite{Ohta2004,Zhang2016},
although
various p-type TCOs such as NiO \cite{NiO1993}, 
Cr$_{2}$O$_{3}$ \cite{CrO1996},
SrCu$_2$O$_{2}$ \cite{SrCu2O21998},
LaCuOS \cite{LaCuOS2000}, 
delafossite series Cu$M$O$_{2}$ \cite{Kawazoe1997,CuScO2000,CuCrMgO2001,CuInO2001,CuYGaO2001,CuBO2007},
V$_2$O$_3$ \cite{VO2019,Zhu2022}, 
and perovskite oxides \cite{Zhang2015,Hu2018}
have been studied so far.
The development of p-type TCOs including computational approach \cite{Brunin2019}
is thus expected for future optoelectronic applications
such as the growing field of the transparent semiconductors in which the p-n junction is essential \cite{Choi2015,Park2020,Hu2020}.

Recently, 
misfit layered cobalt oxide [Ca$_2$CoO$_3$]$_{0.62}$[CoO$_2$] thin films
have been proposed as a peculiar class of p-type efficient TCOs.
As shown in Fig. 1(a), 
the crystal structure of this material is 
two-dimensional, 
consisting of 
alternating rocksalt-type Ca$_2$CoO$_3$ block layers 
and CdI$_{2}$-type triangular-lattice CoO$_2$ conducting layers 
stacking along the $c$ axis \cite{Masset2000,Miyazaki2002,Shikano2003}. 
Interestingly, these two layers possess 
different lattice parameters along the $b$-axis direction to establish incommensurate nature 
as
is also seen in layered chalcogenide materials \cite{Rouxel1995}.
Thus, $\gamma=0.62$ in [Ca$_2$CoO$_3$]$_{\gamma}$[CoO$_2$] is approximate value
for the incommensurate structure, 
and
this material is often called Ca$_3$Co$_4$O$_9$ as the approximate chemical formula.
[Ca$_2$CoO$_3$]$_{0.62}$[CoO$_2$] has been intensively studied as 
an promising oxide thermoelectrics 
because of its large positive Seebeck coefficient of $S \approx 130$~$\mu$V/K near room temperature
owing to the entangled spin and orbital entropy of correlated Co 3d electrons,
coexisting with 
the metallic resistivity \cite{Koshibae2000,Klie2012,Hejtmanek2015,Hebert2016,Yin2023}.
 
For the promising aspect of [Ca$_2$CoO$_3$]$_{0.62}$[CoO$_2$] for TCOs,
Aksit {\it et al} and Fu {\it et al} have reported the optical transmittance and electrical transport measurements 
on the transparent polycrystalline \cite{Aksit2014} 
and $c$-axis oriented \cite{Fu2015} thin films, respectively.
The figure of merit (FOM) for TCOs has been evaluated as
$-1/(R_{\rm sheet} \times \ln T_{\rm opt})$, 
where $R_{{\rm sheet}}$ is the sheet resistance and 
$T_{\rm opt}$ is the transmittance averaged for the specific photon energies in the visible range 
\cite{FOMreference2011,Hu2018}.
Although the evaluated values of FOM
(151~M$\Omega^{-1}$ for the polycrystalline film \cite{Aksit2014} and 
988~M$\Omega^{-1}$ for the $c$-axis oriented film \cite{Fu2015})
are 
still less efficient, 
these previous studies suggest a potential layered optoelectronic oxides 
for the p-type TCOs.

In this paper, 
we attempt to fabricate transparent low-layered [Ca$_2$CoO$_3$]$_{0.62}$[CoO$_2$] single crystals
by utilizing a mechanical tape-peeling method.
This simple peeling method may have a certain merit for the layered crystals 
to keep high carriers mobility \cite{graphene2004}.
In this method, we have fabricated the transparent low-layered single crystals
on the glass substrate with typical thickness of several orders of hundred nanometers.
Through the transmittance and sheet resistance measurements,
we find that the FOM of
the present low-layered crystals is higher than the values reported in the previous studies.
These results indicate that the present low-layered [Ca$_2$CoO$_3$]$_{0.62}$[CoO$_2$] is not only a potential p-type TCO
but also a unique class of transparent thermoelectric oxides with multi-functional properties.

\begin{figure*}[t]
\begin{center}
\includegraphics[width=13cm]{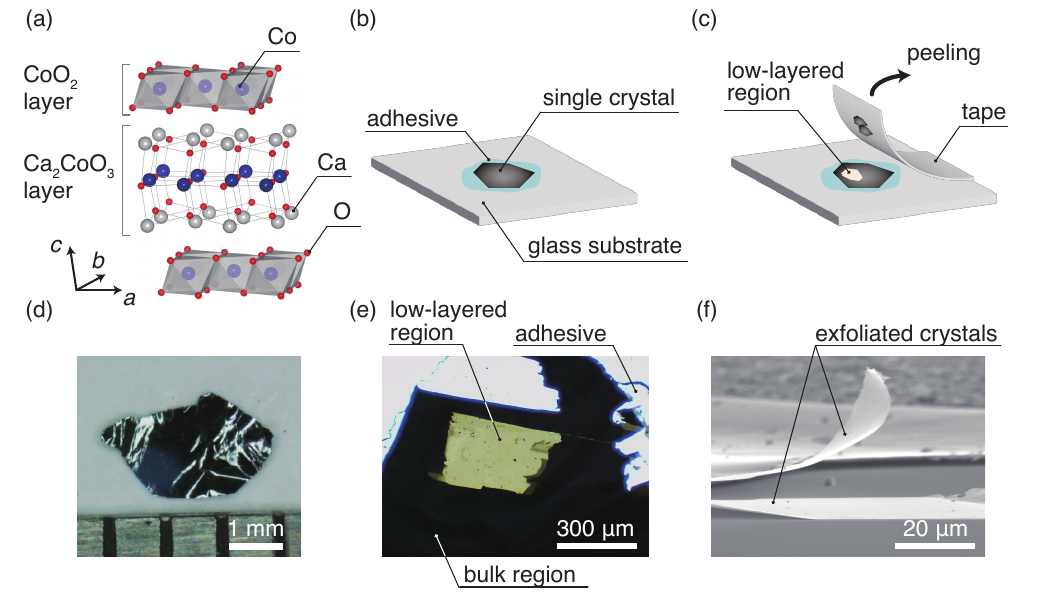}
\caption{
Fabrication process of the low-layered [Ca$_2$CoO$_3$]$_{0.62}$[CoO$_2$] single crystals.
(a) Crystal structure of [Ca$_2$CoO$_3$]$_{0.62}$[CoO$_2$] drawn by VESTA \cite{Momma2011}.
(b) Schematic diagram of the fabrication process. 
Thin crystal is fixed on a glass substrate by using an adhesive bond. 
(c) The low-layered crystal is partly obtained by a tape-peeling method.
(d) Photograph of the bulk single crystal grown by the flux method.
(e) Photograph of the mechanically-exfoliated single crystal. 
Transparent yellow-colored area is the mechanically-exfoliated low-layered region of the crystal.
(f) Side-view SEM image for the small pieces of the mechanically-exfoliated single crystal.
}
\end{center}
\end{figure*}

We have first grown single-crystalline samples of
[Ca$_2$CoO$_3$]$_{0.62}$[CoO$_2$] 
by using a flux method \cite{Mikami2006,Ikeda2016,Saito2017}. 
The low-layered single crystals were fabricated by utilizing a tape peeling method.
The single crystal was first sandwiched by the peeling tapes 
to remove the irregular surfaces of the crystal.
Then, 
as illustrated in Figs. 1(b) and 1(c),
the cleaved crystal was fixed on the glass substrate by using a small amount of transparent adhesive bond 
and mechanically exfoliated by the tape peeling method.
The transparent low-layered region was partly obtained in the single crystal as shown in Fig. 1(e).
Figure 1(f) displays a side-view scanning-electron-microscope (SEM) image for 
small pieces of the exfoliated crystals,
indicating that the crystal thickness is approximately less than 1~$\mu$m.
The present fabrication of low-layered crystals is interesting because
the layered coupling in [Ca$_2$CoO$_3$]$_{0.62}$[CoO$_2$] may have rather strong covalent nature
in contrast to the weak interlayer coupling of van der Waals materials.
Owing to the incommensurate layered structure  of [Ca$_2$CoO$_3$]$_{0.62}$[CoO$_2$],
the samples may become more exfoliative to realize such 
transparent low-layered crystals.
Note that similar layered cobalt oxide 
Bi$_2$Sr$_2$Co$_2$O$_8$ could be exfoliated into nanosheets due to weak van der Waals interlayer coupling \cite{Yu2012,Wang2014,Li2017}.

The transmittance spectra were measured 
by using a USB module of a complementary metal-oxide semiconductor (CMOS) spectrometer equipped with an optical microscope. 
We evaluated the transmittance $T(\omega)$ as $T(\omega) = I_{\rm sample}(\omega)/I_{\rm ref}(\omega)$, 
where $I_{\rm sample}(\omega)$ is the intensity spectrum for the low-layered crystal on the glass substrate 
and 
$I_{\rm ref}(\omega)$ is the reference spectrum for the glass substrate without samples.
To consider the reflection at the interfaces, the absorbance was obtained as follows:
The measured intensity $I_{\rm sample}$ is given as 
$I_{\rm sample}=I_0(1-R_1)(1-R_2)e^{-\alpha d}$,
where $I_0$ is the intensity of incident light, 
$R_1$ and $R_2$ are the reflectance at the crystal-glass and the crystal-air interfaces, respectively,
and 
$\alpha$ is the absorption coefficient of the crystal. 
Note that the multiple reflection is negligibly small.
Also, the reference intensity $I_{\rm ref}$ is given as $I_{\rm ref}=I_0(1-R_3)$,
where $R_3$ is the reflectance at the glass-air interface.
The measured transmittance $T$ is then given as 
$T = I_{\rm sample}/I_{\rm ref} = e^{-\alpha d}/\gamma$,
where $\gamma = \frac{1-R_3}{(1-R_1)(1-R_2)}$.
According to Fresnel’s equation, 
the interface reflection between the media $\alpha$ and $\beta$ is calculated as 
$|(\tilde{n}_{\alpha}-\tilde{n}_{\beta})/(\tilde{n}_{\alpha}+\tilde{n}_{\beta})|^2$,
where $\tilde{n}_{\alpha}$ and $\tilde{n}_{\beta}$ are 
the complex refraction indices of the media $\alpha$ and $\beta$, respectively.
We used $\tilde{n}_{\rm air}=1.0$ and $\tilde{n}_{\rm glass}=1.5$. 
The complex refraction index of the crystal $\tilde{n}_{\rm crystal}=n(\omega)+i\kappa(\omega)$ is obtained from the Kramers-Kronig (KK) analysis \cite{Tanabe2016},
where $n(\omega)$ is the refractive index and $\kappa(\omega)$ is the extinction coefficient.
The absorbance of the crystal is then calculated from the measured transmittance $T$ as
$A_{\rm exp} = \alpha d\log_{10}e = -\log_{10}(\gamma T)$.

The sheet resistance $R_{\rm sheet}$ of the low-layered crystals was measured by using a conventional van der Pauw method.
The excitation current of $I=10$~$\mu$A was applied by Keithley 6221 current source and 
the voltage was measured with synchronized Keithley 2182A nanovoltmeter. 
These two instruments were operated in a built-in Delta mode to cancel the offset voltage.
We used a closed refrigerator to evaluate the temperature dependence of the sheet resistance below room temperature.

\begin{figure}[t!]
\begin{center}
\includegraphics[width=7cm]{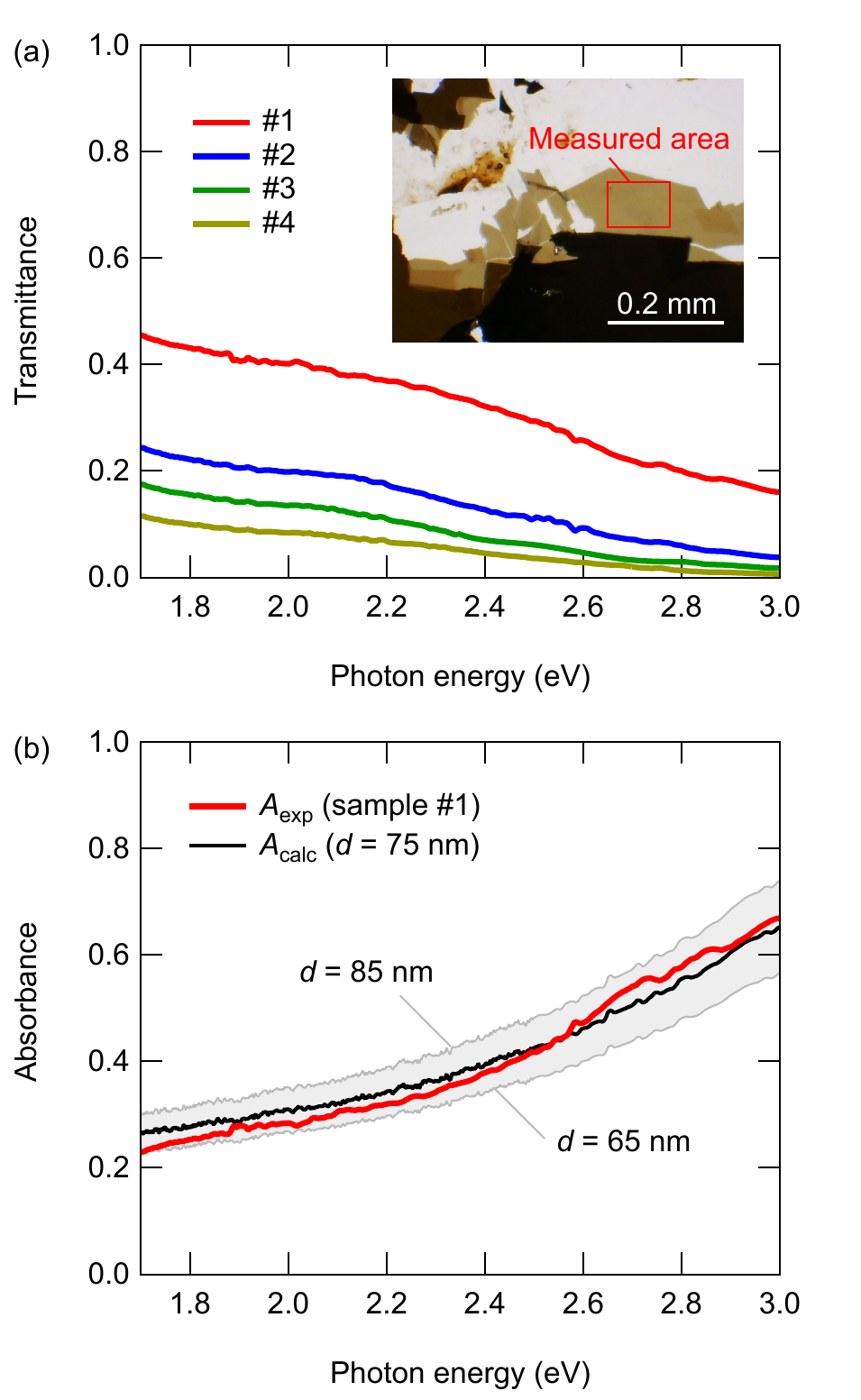}
\caption{
(a) Transmittance spectra of the low-layered [Ca$_2$CoO$_3$]$_{0.62}$[CoO$_2$] single crystals. 
The inset shows the photograph of the measure crystal and 
the red rectangular indicates the measured area for the transmittance spectra.
(b) Absorbance spectra. In the present study, 
we compare the absorbance spectra obtained from the present transmittance measurements, $A_{\rm exp}$,
with the calculated data $A_{\rm calc}$ from the earlier optical study by assuming several values of the sample thickness $d$.
See text for details.
}
\end{center}
\end{figure}

Figure 2(a) depicts the transmittance spectra $T(\omega)$ of the low-layered [Ca$_2$CoO$_3$]$_{0.62}$[CoO$_2$] single crystals
in the visible range.
The inset of Fig. 2(a) shows the optical image of the measured sample and the measured area is indicated by the red rectangular.
The measured low-layered region is yellow-colored transparent and the averaged transmittance in the visible range is about 0.3,
indicating that thin low-layered crystal is fabricated in the present peeling process using single crystals.

To estimate the thickness of the low-layered crystals, 
we now consider the absorbance of the crystal.
As described before,
we evaluate the absorbance spectra of the low-layered crystals $A_{\rm exp}(\omega)$ from the measured transmittance $T(\omega)$
The absorbance spectra of typical low-layered crystal $A_{\rm exp}(\omega)$ is shown in Fig. 2(b) with the red curve.
Then, we also
calculate the absorbance spectra $A_{\rm calc}(\omega)$ from the reflectivity data in the previous study \cite{Tanabe2016}.
Using the extinction coefficient  $\kappa(\omega)$ obtained by the KK analysis,
we
calculate $A_{\rm calc}(\omega)$ for several values of the sample thickness $d$ as
\begin{align}
A_{\rm calc}(\omega) = \frac{1}{\ln{10}}\frac{2\omega \kappa(\omega)}{c}d,
\end{align}
where $c$ is the speed of light.
In Fig. 2(b), 
we also plot $A_{\rm calc}(\omega)$ for several values of $d$,
and 
$A_{\rm calc}(\omega)$ data calculated with $d = 75$~nm is matched with 
$A_{\rm exp}(\omega)$ data measured in the present study, indicating that
the thickness of the present low-layered crystal is $d \approx 75$~nm and the uncertainty is approximately 10 nm.
We evaluate the thickness of several crystals in the same manner, 
and
typical value of the thickness is found to be a few hundred nanometers,
consistent with the 
SEM image for the side view of low-layered crystals shown in Fig. 1(f).

\begin{table*}[t!]
  \centering
  \begin{tabular}{cccccccc} \hline
  Samples & No.& $d$ (nm) &$R_{\rm sheet}$ ($\Omega/\square$) &$\rho$ (m$\Omega$cm) & $T_{\rm opt}$ ($\%$)  &FOM (M$\Omega^{-1}$) & References \\ \hline
  \multirow{4}{*}{Low-layered single crystals}    &\#1 & {75(10)} & 556 & {4.2(6)} & {31.1} & {1540} &\multirow{4}{*}{Present study}  \\ \cline{2-7}
   &\#2 &  {150(10)}& 304 & {4.6(3)} & {13.4} & {1637} &  \\ \cline{2-7}
    &\#3& {190(15)} & 227 & {4.3(3)} & {8.4} & {1779} & \\ \cline{2-7}
  &\#4  & {240(15)} & 176 & {4.2(3)} & {5.2} & {1922} & \\ \hline
 Polycrystalline film &  & 100 & 5700 & 57 & 31.3 &    151   &  \cite{Aksit2014} \\ \hline
 {$c$-axis oriented} film&   & 50 &  1460  & 7.3 &50 & 988   &  \cite{Fu2015} \\ \hline
  \end{tabular}
  \caption{The room-temperature optoelectronic properties and the figure of merit (FOM) of transparent [Ca$_2$CoO$_3$]$_{0.62}$[CoO$_2$] in various forms.}
  \label{tb:mulrow}
\end{table*}

\begin{figure}[t!]
\begin{center}
\includegraphics[width=8cm]{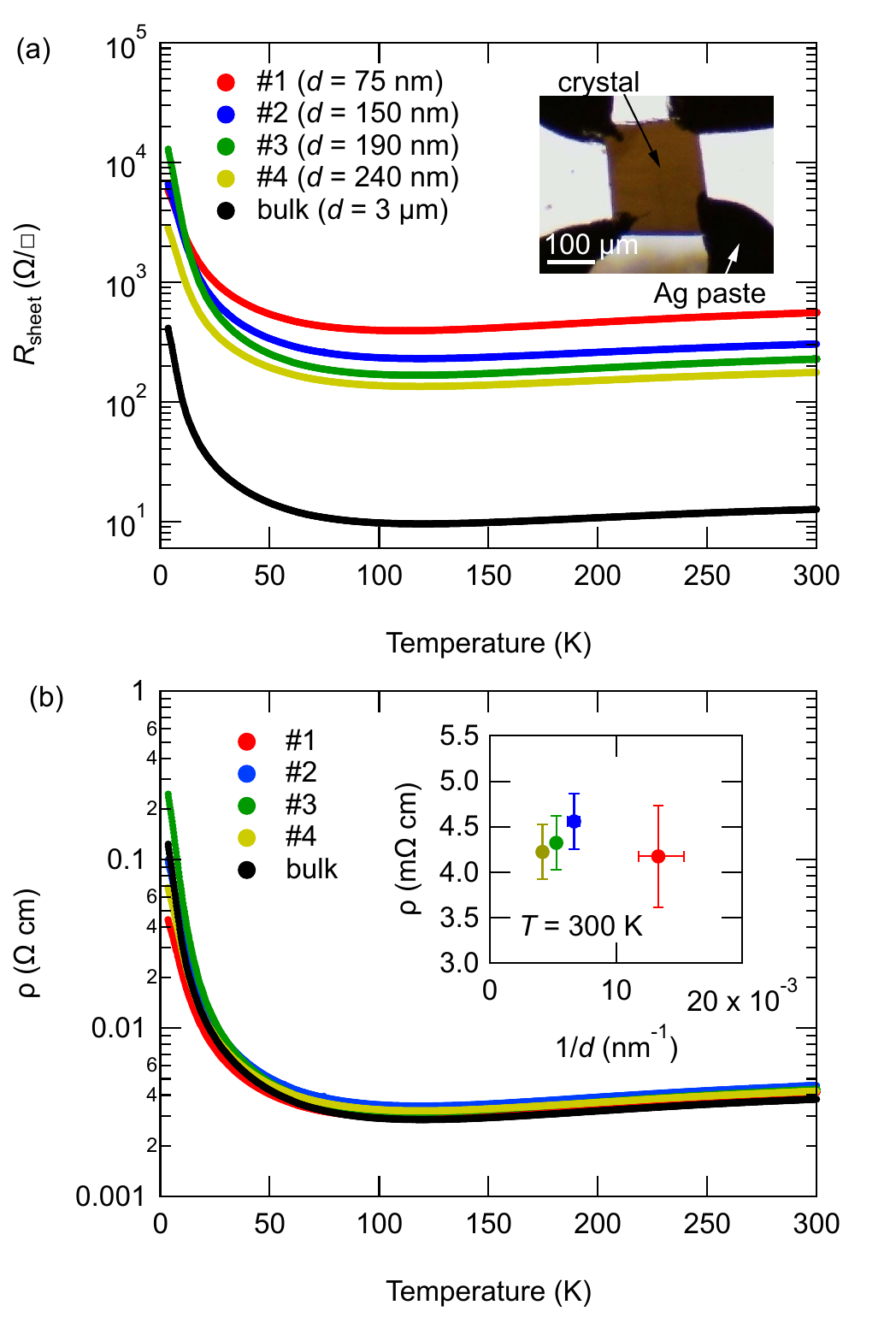}
\caption{
(a) Temperature dependence of the sheet resistance $R_{\rm {sheet}}$ 
obtained by the van der Pauw method. 
The inset shows a photograph of the low-layered rectangular-shaped 
sample attached with four electrical contacts at the corners.
(b) Temperature dependence of the resistivity $\rho$ evaluated as $\rho=R_{\rm sheet} \times d$.
The inset shows the 
the resistivity measured at 300 K as a function of the inverse of the thickness, $1/d$,
for four low-layered crystals.
}
\end{center}
\end{figure}

To examine the sheet resistance of the low-layered transparent single crystals,
we have prepared a rectangular-shaped low-layered sample by cutting the exfoliated crystal
and made four electrical contacts at the corners by using a silver paint 
as shown in the inset of Fig. 3(a).
The sheet resistance $R_{\rm {sheet}}$ is then measured by the conventional van der Pauw (vdP) method,
and the obtained temperature dependence of $R_{\rm {sheet}}$ and 
the resistivity $\rho=R_{\rm sheet} \times d$ are shown in Figs. 3(a) and 3(b), respectively.
Note that
the measured resistivity in the vdP configuration
is given as
the geometric mean of $\rho = \sqrt{\rho_{a}\rho_{b}}$,
where $\rho_i$ $(i=a,b)$ is the resistivity measured along $i$ direction \cite{vdp},
since
this material exhibits 
slight in-plane transport anisotropy \cite{Sakabayashi2021}.

Overall behavior of the temperature dependence of the resistivity in the present low-layered 
crystals agrees well with the previous data \cite{Murashige2017,Sugiura2007};
near room temperature, a metallic resistivity of $\rho \approx 4$~m$\Omega$ cm
is achieved and a semiconducting temperature dependence is observed below 100~K owing to 
the carrier localization effect \cite{Bhaskar2014}
or a pseudo-gap formation associated with the spin-density-wave ordering \cite{Hsieh2014}.
Thus, these results indicate that 
a single-crystalline transparent conducting oxide is well fabricated in 
the layered [Ca$_2$CoO$_3$]$_{0.62}$[CoO$_2$] with no significant damage in the peeling process.
Note that the low-temperature behavior seems sample-dependent, which probably comes from 
extrinsic effects such as the oxygen defects.

Generally, the resistivity of metallic films depends on the thickness 
due to the surface scattering as
discussed in the Fuchs-Sondheimer model \cite{Chawla2011}.
In such case, the resistivity may show a linear dependence of 
$\rho = \rho_0(1+\Lambda/d)$,
where $\rho_0$ and $\Lambda$ are the bulk-crystal resistivity and a characteristic length, respectively.
The characteristic length $\Lambda$ is comparable to a mean free path of carriers near surfaces.
We then plot the resistivity measured at 300 K as a function of the inverse of the thickness, $1/d$,
for the low-layered crystals in the inset of Fig. 3(b), and 
find no significant thickness dependence.
This result
is consistent with the fact that 
the mean free path of [Ca$_2$CoO$_3$]$_{0.62}$[CoO$_2$] is about 1~nm \cite{Hsieh2014},
which is much smaller than the thickness of the  low-layered crystals,
and indicates that the surface scattering is not dominant in the present case.
On the other hand,
a thickness-dependent metal-insulator transition has been observed in the nanosheets Bi$_2$Sr$_2$Co$_2$O$_8$ \cite{Wang2014},
implying a strong thickness dependence of the resistivity in the layered oxides.
Also, interesting spin-related surface transport \cite{Huang2017} and 
two-dimensional quantum confinement effect \cite{Zhou2018}
have been suggested for the resistive behavior in the metallic films.
A more detailed consideration may be required as a future study
along with further magneto-transport experiments on the thinner low-layered crystals.

\begin{figure}[t!]
\begin{center}
\includegraphics[width=8cm]{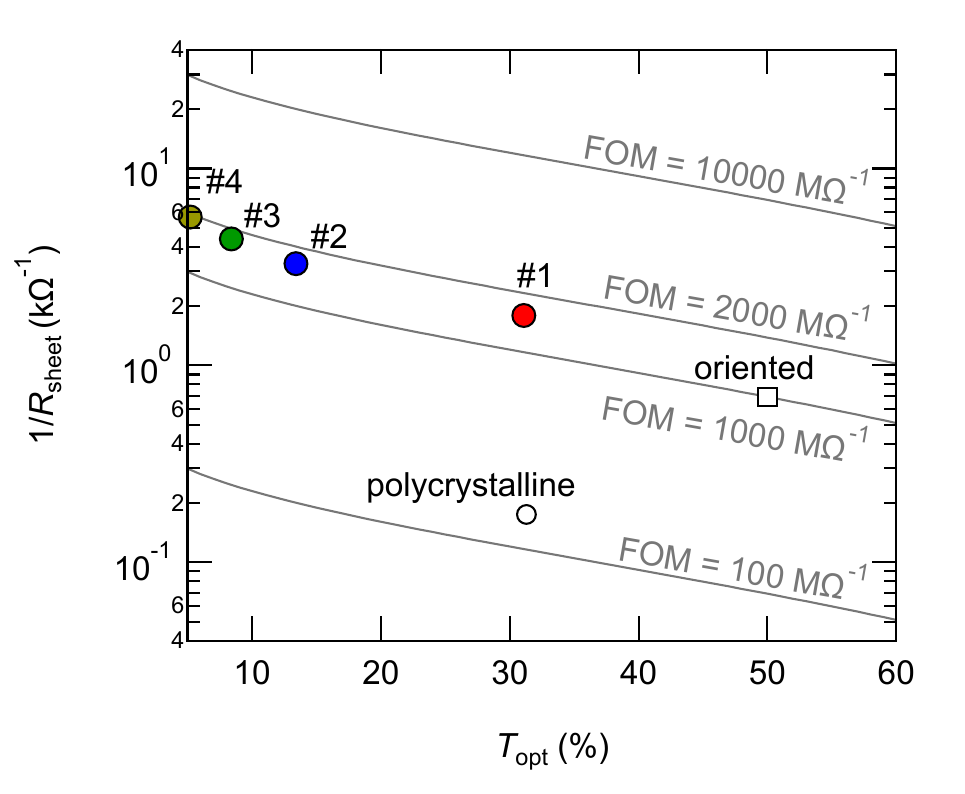}
\caption{
Single logarithmic plot of 
the inverse sheet resistance $1/R_{\rm sheet}$ vs 
the transmittance averaged for specific photon energies $T_{\rm opt}$
for various forms of [Ca$_2$CoO$_3$]$_{0.62}$[CoO$_2$].
The data for the polycrystalline and the $c$-axis oriented films are taken from Refs. \onlinecite{Aksit2014,Fu2015},
respectively.
Solid curves represent the calculated curves for several values of FOM.
}
\end{center}
\end{figure}

Having the observed transmittance and the sheet resistance, 
we now discuss the figure of merit (FOM) of the low-layered [Ca$_2$CoO$_3$]$_{0.62}$[CoO$_2$] single crystals.
The FOM for TCOs has been evaluated with two different procedures \cite{Hu2018}:
one is proposed by Haacke as FOM$^{\rm H}=T_{\rm opt}^{10}/R_{\rm sheet}$,
where $R_{\rm sheet}$ is the sheet resistance and 
$T_{\rm opt}$ is the transmittance averaged for specific photon energies
($\hbar\omega = 1.77, 2.0, 2.25, 2.5, 2.75, 3.0$~eV).
This evaluation however tends to overweight the importance of the transparency.
The other is widely used one proposed by Gordon as
\begin{align}
{\rm FOM}^{\rm G} = -\frac{1}{R_{\rm sheet} \times \ln T_{\rm opt}},
\end{align}
which is also expressed as 
${\rm FOM}^{\rm G} = \sigma/\alpha$
by using
the electrical conductivity $\sigma = 1/(R_{\rm sheet}d)$ and the 
absorption coefficient $\alpha \approx -(1/d)\ln T_{\rm opt}$.
Since FOM$^{\rm G}$ is evaluated in the earlier works on this material,
we also use FOM$^{\rm G}$ as the figure of merit in this paper.

Table I summarizes the optoelectronic properties and 
the FOM of transparent [Ca$_2$CoO$_3$]$_{0.62}$[CoO$_2$] in 
the present study and previous reports \cite{Aksit2014,Fu2015}.
Owing to the single-crystalline nature, 
the resistivity is lower than those in the previous reports \cite{Aksit2014,Fu2015,Yin2021},
leading to large values of FOM.
Note that the resistivity of the high-quality epitaxial films is comparable to that in the low-layered crystals \cite{Sugiura2007}
while 
the transmittance of the epitaxial films has not been evaluated.
We also note that the FOM values slightly decrease with decreasing thickness:
As mentioned previously, 
the measured transmittance is $T =e^{-\alpha d}/\gamma$.
Thus, using the conductivity $\sigma = 1/(R_{\rm sheet}d)$, 
the FOM is given as 
${\rm FOM} = \sigma /(\alpha  + \frac{1}{d}\ln \gamma)$.
Since $\sigma$ and $\alpha$ are thickness-independent parameters in principle and 
$\gamma$ is larger than unity,
this equation may indicate that the FOM value decreases with decreasing $d$
owing to the reflectance at the interfaces in the present devices.
Nonetheless, the FOM value in the present study 
is the same order of those in high-performance $p$-type TCOs such as
CuCr$_{1-x}$Mg$_x$O$_2$ (FOM $\sim 5000$~M$\Omega^{-1}$) \cite{CuCrMgO2001}
and La$_{2/3}$Sr$_{1/3}$VO$_{3}$ (FOM $\sim 7000$~M$\Omega^{-1}$) \cite{Hu2018}.
We note that doping effect such as Bi substitution \cite{Hsieh2014} may be promising
to further enhance the FOM in the low-layered [Ca$_2$CoO$_3$]$_{0.62}$[CoO$_2$].
Figure 4 represents the single logarithmic plot of the sheet conductance $1/R_{\rm sheet}$ 
as a function of $T_{\rm opt}$.
In Fig. 4, we plot the room-temperature data for four low-layered single crystals as well as the data
from an earlier study of the polycrystalline and the $c$-axis oriented films \cite{Aksit2014,Fu2015},
along with the calculated curves for several values of FOM.
Indeed, the present low-layered crystals exhibit higher value of FOM compared to 
those in previous study, indicating that
the present peeling technique using single crystals is  efficient certainly to realize high-performance TCOs.

To summarize,
we have fabricated the transparent low-layered [Ca$_2$CoO$_3$]$_{0.62}$[CoO$_2$] single crystals
on the glass substrate
by utilizing mechanical peeling method.
From the results of the optical and transport measurements,
we find that 
the figure of merit 
in the present crystals is higher than the values reported in the previous studies on
the polycrystalline and $c$-axis oriented films
owing to the lower resistivity in the single-crystalline samples.
Since [Ca$_2$CoO$_3$]$_{0.62}$[CoO$_2$] has been studied as a potential oxide thermoelectrics,
the present study offers
a unique class of transparent thermoelectric oxides with multi-functional properties \cite{Ishibe2023,Darnige2023}.
Also, the present low-layered crystals belonging to the strongly correlated electrons system may
provide a fascinating playground 
to investigate the nature of
correlated electrons confined in low dimension.

\subsection*{Conflict of Interest}

The authors have no conflicts to disclose.

\section*{Data Availability}

The data that support the findings of this study are available from the corresponding author (R. Okazaki) upon reasonable request.

\section*{Acknowledgments}

We appreciate K. Tanabe for providing us with the reflectivity data of [Ca$_2$CoO$_3$]$_{0.62}$[CoO$_2$] single crystals.
This work was supported by JSPS KAKENHI Grant No. 22H01166 and Research Foundation for the Electrotechnology of Chubu (REFEC).


\end{document}